# Effect of Chemical Structure on the Isobaric and Isochoric Fragility in Polychlorinated Biphenyls

C.M. Roland[1] and R. Casalini[2]
Naval Research Laboratory, Chemistry Division, Code 6120, Washington DC 20375-5342

**Abstract**

Pressure-volume-temperature data, along with dielectric relaxation measurements, are reported for a series of polychlorinated biphenyls (PCB), differing in the number of chlorine atoms on their phenyl rings. Analysis of the results reveals that with increasing chlorine content, the relaxation times of the PCB become governed to a greater degree by density, $\rho$, relative to the effect of temperature, T. This result is consistent with the respective magnitudes of the scaling exponent, $\gamma$, yielding superpositioning of the relaxation times measured at various temperatures and pressures, when plotted versus $\rho^{\gamma}/T$. While at constant (atmospheric) pressure, fragilities for the various PCB are equivalent, the fragility at constant volume varies inversely with chlorine content. Evidently, the presence of bulkier chlorine atoms on the phenyl rings magnifies the effect density has on the relaxation dynamics.

[1] roland@nrl.navy.mil     [2] casalini@ccf.nrl.navy.mil

**Introduction**

Relating the dynamics of molecules to their chemical structure is of obvious fundamental interest, and a necessary step in understanding the origin of the macroscopic physical properties. Among the various relaxation properties, most intriguing are those associated with the supercooled regime just above the glass transition temperature, $T_g$. Complex behaviors become apparent, including decoupling of translational motions from the reorientational dynamics[1,2,3], a change in the temperature-dependence of the dynamic properties at some temperature $T_B > T_g$ [4,5,6,7,8,9] which moreover occurs at a material-characteristic value of the relaxation time [6,10], and the splitting off from the glass transition of a higher frequency secondary relaxation or an "excess wing" phenomenon.[11,12,13,14,15] At higher frequencies, or observed below $T_g$, is a broad span of a nearly constant loss in the susceptibility [16,17,18,19,20], followed by the Boson peak and vibrational motions [21,22,23,24,25,26]. How these various phenomena relate to structural relaxation and $T_g$ is a central issue in condensed matter physics.

In this work we explore the connection between the chemical structure of polychlorinated biphenyls (PCB) and their local dynamics. PCB are inert, thermally stable liquids, comprised of various isomers. They are readily cooled or compressed without crystallization, and undergo a glass transition at a temperature dependent on their chlorine content. Many investigations of PCB have been reported, with early work exploiting their solvent power. Experiments on dilute PCB solutions were the first to show that dissolved polymer chains modify the local motions of the solvent molecules[27,28,29,30]. A very anomalous form of this modification was observed in mixtures of PCB with polybutadiene. Usually, the glass transition temperature of a mixture, as well as its relaxation times, is intermediate between those of the neat components. However, the addition of lower $T_g$ polybutadiene *decreases* the PCB relaxation time[31,32,33,34]. This interesting anomaly was subsequently seen in PCB/polystyrene mixtures[35,36], and more generally in both polymer solutions[37,38] and blends[39].

In the last decade, the availability of PCB has been severely limited, with the bulk of research directed to environmental and toxicological issues. Nevertheless, the facile glass-forming ability of the liquids offers an opportunity to investigate structure-property relationships. Both the fragility ($T_g$-normalized temperature dependence of the relaxation times, τ) and the shape of the relaxation function of PCB are independent of chlorine content[40]. The relaxation function follows the Kohlrausch-Williams-Watts (KWW) form, with a stretch exponent equal to

~0.65. However, as measured by dielectric spectroscopy, there is a deviation from the KWW function on the high frequency side of the structural relaxation peak. This so-called "excess wing" (EW) becomes less prominent with increasing Cl content[40]. When compared at a fixed value of the relaxation time, the shape of both the main peak and the EW is constant; that is, they depend on τ(T,P), but not on the particular values of T and P[41]. At atmospheric pressure a change in the dynamics is observed, corresponding to a change in τ(T), at a temperature $T_B = 1.14\ T_g$.[40] While this $T_B$ increases with pressure, the value of the relaxation time at $T_B$ is invariant to pressure (and also to chlorine content).[42,43]

To unify the dynamic behavior of glass-forming liquids and polymers, it is of interest to obtain an analytical form for the relaxation times, which explicitly quantifies the respective dependences of τ on temperature and density. Efforts toward this end are based on some model for the glass-transition process, such as free volume [44,45,46,47] or thermal activation [48,49,50,51]. One approach is to regard structural relaxation as an activated process, with a density-dependent activation energy; thus, τ becomes a function of E(ρ)/T [52,53,54]. A Lennard-Jones 6-12 (LJ 6-12) intermolecular potential, in which the local dynamics are dominated by the repulsive term, suggests a $\rho^4/T$ form for the temperature and density dependences of local processes. And indeed, for a glass-former such as o-terphenyl, in which intermolecular interactions can be accurately described by a LJ 6-12 potential[55,56], relaxation times measured by neutron [52] and light scattering [53] at various T and P fall on a single curve when plotted versus $\rho^4/T$.

We have generalized this idea, to show that over a wide range of temperatures, encompassing even the change in dynamics at $T_B$, dielectric relaxation times for many glass-forming liquids [57,58] and polymers [59,60] can be expressed as a single function of $\rho^\gamma/T$ [61], in which γ is a material specific constant, whose magnitude depends on the degree to which density governs the relaxation times. The extremes cases, γ = 0 and ∞, correspond respectively to purely temperature-driven dynamics (e.g., limiting behavior at high temperatures) and hard spheres. The magnitude of this exponent can be plausibly related to the intermolecular repulsive potential. [52,62,63,64,65] More recently, a dynamic light scattering study of various glass-formers [66] found similar $\rho^\gamma/T$ scaling of the relaxation times, with values γ equivalent to those measured dielectrically.

Since τ only depends on $T\rho^{-\gamma}$, it can be shown that the ratio of the isochoric and isobaric activation enthalpies, $E_V/E_P$ ( $E_X = 2.303RT \left. \frac{d\log\tau}{dT^{-1}} \right|_X$ ) at $T_g$ varies with γ according to [58]

$$E_V/E_P \big|_{T_g} = (1 + T_g \alpha_P(T_g) \gamma)^{-1} \qquad (1)$$

where $\alpha_P$ is the thermal expansion coefficient at atmospheric pressure. This activation enthalpy ratio is of great significance, since it provides a direct measure of the degree to which changes in τ with temperature result from the accompanying volume changes, as opposed to the changes in thermal energy.[67,68]

The temperature dependence of glass-formers is often characterized by their isobaric fragility, $m_P \equiv \left. \frac{d\log\tau(T_g)}{d(T_g/T)} \right|_P$, where $m_P$ can range from ~15 for orientationally disordered crystalline materials to almost 200 for polymers [69]. Much effort has been expended in correlating fragility with other dynamic properties and with the thermodynamics, in order to identify the general principles underlying vitrification. The magnitude of the fragility has been related to (i) the breadth of the relaxation function [69,70], (ii) the Debye-Waller factor [71], (iii) the T-dependence of the configurational entropy[72], (iv) the liquid shear modulus[73] or its value relative to the bulk modulus[74], (v) vibrational properties of the glass[22], and (vi) the form of the interaction potential[50,75,76]. Since the fragility of PCB at atmospheric pressure is independent of chemical structure (i.e., Cl content) [40], we expect similarities in their various dynamic and thermodynamic properties, at least to the extent correlations of the latter with $m_P$ are valid.

In this work, we report PVT measurements on three PCB, and combine these data with published and new dielectric relaxation measurements. The results enable a systematic analysis of the relation of chemical structure to the relaxation properties in the supercooled state. We show that, notwithstanding the equivalence of their isobaric atmospheric-pressure fragilities and other dynamical properties (stretch exponent, $T_B$, etc.), the PCB exhibit marked differences in their dynamics. As a metric for glass transition behavior, the isobaric fragility has limitations, due to its lack of consideration of density effects. Changes due to thermal energy and to density are convoluted in isobaric measurements, so that an unambiguous understanding is possible only if one of these variables dominates. However, this is rarely the case, the exception being strongly associated glass-formers, for which temperature may be the dominant control variable.[68,77,78] We determine herein the isochoric (constant density) values of fragility, $m_\rho \equiv \left.\dfrac{d\log\tau(T_g)}{d(T_g/T)}\right|_\rho$, and show how this quantity, in combination with the $\rho^\gamma/T$ scaling of $\tau$ described above, provides a clearer delineation of the factors governing the supercooled dynamics.

**Experimental**

The polychlorinated biphenyls (Monsanto Aroclors obtained from J. Schrag of the University of Wisconsin), were PCB42, PCB54, and PCB62, where the number refers to the average chlorine content by weight.

Dielectric measurements were carried out using a parallel plate geometry with an IMASS time domain dielectric analyzer ($10^{-4}$ to $10^4$ Hz) and a Novocontrol Alpha Analyzer ($10^{-2}$ to $10^6$ Hz). For measurements at elevated pressure, the sample was contained in a Manganin cell (Harwood Engineering), with pressure applied using a hydraulic pump (Enerpac) in combination with a pressure intensifier (Harwood Engineering). Pressures were measured with a Sensotec tensometric transducer (resolution = 150 kPa). The sample assembly was contained in a Tenney Jr. temperature chamber, with T variations at the sample less than 0.1 K.

PVT experiments employed a Gnomix apparatus[79], modified to allow measurements at sub-ambient temperatures. Changes in volume of the liquid PCB were determined isothermally at pressures from 10 to 200 MPa, over a temperature range from as low as -15 up to 130°C. The data were converted to specific volumes, V (= $1/\rho$), using the value of V measured for ambient conditions with a pycnometer.

**Results**

Representative PVT data are shown in Fig. 1 for PCB54. The pressure increments were 10 MPa, yielding *ca.* 600 V(T,P) data points. At the pressure-dependent $T_g$, there is an increase in the thermal expansivity. For temperatures above this, specific volumes for the (equilibrium) liquid can be represented using the Tait equation of state[80]

$$V(T,P) = (a_0 + a_1 T + a_2 T^2)\left[1 - 0.0894\ln\left(1 + P/b_0\exp(b_1 T)\right)\right] \qquad (2)$$

The fit parameters, $a_0$, $a_1$, $a_2$, $b_0$, and $b_1$, for the three PCB samples are listed in Table 1.

Dielectric relaxation times for PCB54, defined as the inverse of the frequency of the maximum in the dielectric loss, are shown as a function of pressure in Fig. 2. The usual measure of pressure dependence is the activation volume, $\Delta V = RT \left.\frac{\partial \ln \tau}{\partial P}\right|_T$. This presumes a linear relationship between ln $\tau$ and P, which is accurate for the PCB54 up through the highest pressures (332 MPa) at the four temperatures in Fig. 2. The obtained activation volumes, displayed in the inset to the figure, show the expected decrease with temperature, $\frac{d\Delta V}{dT} = -1.28$ ($\pm 0.05$) mLmol$^{-1}$K$^{-1}$.

Defining a dynamic glass transition as the temperature at which $\tau = 10$ s,[81] the pressure dependence of $T_g$ can be described using the Andersson relation[82]

$$T_g = k_1 (1 + \frac{k_2}{k_3} P)^{1/k_2} \qquad (3)$$

an empirical equation derivable from the Avramov structural relaxation model [83]. Fitting eq. 3 to the PCB54 data, we obtain $k_1 = 250 \pm 3$ K, $k_2 = 2.7 \pm 0.6$, and $k_3 = 780 \pm 120$ MPa; thus, in the limit of zero pressure, $dT_g/dP = 0.30 \pm 0.01$ K/MPa. Results are shown for the three PCB in Fig. 3. The pressure coefficients, tabulated in Table 2, increase with increasing chlorine content, suggestive of an increasing influence of density on $\tau$.

Notwithstanding their differences in $dT_g/dP$, the isobaric fragilities of the PCB, as measured at ambient pressure, are all equal [40]. These data are shown in Figure 4, with $m_P = 59.2 \pm 0.7$ (using $\tau(T_g) = 10$ s). Thus, as temperature is lowered, the reduction in thermal energy and concomitant increase in density have the same net effect on $\tau$ for the three liquids, at least when data for the PCB are compared at equal $T/T_g$. In order to assess directly the consequences of changes in temperature and density, in Figure 5 we plot all relaxation times measured for the three samples as a function of $\rho^\gamma/T$. The exponent $\gamma$ is adjusted, independently for each material, to bring into coincidence the data measured as a function of temperature at fixed (ambient) P and as a function of pressure at various fixed T. Good superpositioning is obtained, with the $\gamma$ values listed in Table 2. They rank order as PCB42 < PCB54 < PCB62. A larger value of this density exponent of course indicates a stronger influence of $\rho$, relative to that of thermal energy, on the T-dependence of $\tau$. This relationship is quantified by eq. 1, from which we calculate the activation enthalpy ratios shown in Table 2. When the change in relaxation times with temperature is due equally to changes in density and thermal energy, $E_V/E_P = 0.5$. Thus, for PCB62, for which $E_V/E_P = 0.38$, density changes exert a stronger effect on $\tau(T)$ than changes in thermal energy.

To corroborate this result, which is based on the superposition of the $\tau(T,P)$, we calculate the ratio of the thermal expansion coefficient for a fixed value of the relaxation time, $\alpha_\tau (= -V^{-1} (\partial V/\partial T)_\tau)$ to its magnitude at constant pressure, $\alpha_P (= -V^{-1} (\partial V/\partial T)_P)$. This ratio, $|\alpha_\tau|/\alpha_P$, will be significantly larger than one if thermal energy, rather than density, governs the variation of $\tau$ with temperature.[84] Using the PVT results in Table 1, along with the relaxation data for the three samples, we calculate the expansivity ratios at $T=T_g$, $P = 0.1$ MPa and $\tau = 10$ s (Table 2). From these, the activation enthalpy ratio is calculated as [85]

$$E_V / E_P = (1 - \alpha_P/\alpha_\tau)^{-1} \tag{4}$$

Eq. 4 yields values equivalent to those determined using eq. 1, corroborating the scaling shown in Fig. 5.

Since the relaxation times are a function only of $\rho^\gamma/T$, we can calculate $\tau$ for any condition of T and P. This enables relaxation times to be obtained for constant density conditions, something not experimentally feasible. We choose the value of density prevailing at the ambient pressure $T_g$, and then determine for each T the P such that $\rho(T,P) = \rho(T_g, 0.1$ MPa$)$. The value of the relaxation times is then obtained using the fact that $\tau$ is uniquely determined by $\rho^\gamma/T$. The results are shown in Fig. 4 as a function of $T_g$-normalized temperature. The slopes of these curves define an isochoric fragilty, $m_\rho$, which for all cases is less than the corresponding isobaric value. A similar result has been found for other glass-formers [86,87,88]. Interestingly, at constant $\rho$ the fragile character of the T-dependence is almost completely removed, such that the isochoric temperature dependence becomes almost Arrhenius. Moreover, while at constant pressure the fragility is the same for the three samples, at constant density it decreases with increasing chlorine content of the PCB. Since $m_\rho$ represents the limiting high-pressure value of the fragility, the pressure coefficient for the three samples is negative, $\frac{dm_P}{dP} < 0$. This is generally in accord with results for other materials [70,88,89,90,91,92].

**Summary**

An intriguing feature of the PCB is that different congeners, having significantly different chlorine content, exhibit the same isobaric fragility (Fig. 4). This would seem to suggest that the relaxation behaviors are the same, apart from the differences in $T_g$. However, the results herein make clear that the supercooled dynamics of these three liquids are quite distinct. Increasing chlorine content results in a systematically stronger influence of $\rho$ on $\tau(T)$. This is seen directly in the values of the activation enthalpy ratios (Table 2). The role of density is also inferred from the superpositioning of the relaxation times in Fig. 5 – a larger scaling exponent (stronger $\rho$ effect) is associated with PCB having more chlorine atoms on the phenyl rings. Thus, in materials having the same type of molecular structure, $T_g$ can be controlled by changing the intermolecular repulsion (in the present case, by altering the number of chlorine atoms). By weakening this repulsion (smaller $\gamma$), thermal energy becomes more dominant, whereby the glass transition is reached at lower temperatures. Conversely, by making volume more dominant (through less flexible bonds or the introduction of bulky pendant groups), $T_g$ increases. Note that the isobaric fragility is expected to be enhanced by the presence of pendant groups, based on the idea that steric hindrances enhance intermolecular cooperatitvity[93]. However, the analysis herein shows quantitatively that this is a direct effect of density.

The limitation of characterizing relaxation properties using the conventional isobaric fragility is its failure to distinguish the influence of the energy landscape on the dynamics from density effects *per se* (although, of course, density affects the landscape). The contributions of temperature and density must be separately quantified, in order to understand what governs the dynamics of glass-formers. In this regard, it is tempting to draw a connection between the magnitude of the scaling exponent and the nature of the intermolecular potential.[63,64,65] For all three PCB, $\gamma$ is larger than the value (= 4) for a LJ 6-12 fluid, indicating a fairly hard potential. A strong distance-dependence justifies the assumption, implicit in the scaling approach, that the attractive interactions can be neglected for local properties; that is, they are manifested only as a background pressure. However, this is inappropriate for global properties, wherein the details of intermolecular interactions become important. For this reason, the equation of state cannot be expressed in terms of $\rho^\gamma/T$, notwithstanding the dependence of $\tau$ on this same variable.

**Acknowledgment**

This work was supported by the Office of Naval Research.

**Table 1** Equation of state parameters for PCB above $T_g$

|  | PCB42 | PCB54 | PCB62 |
|---|---|---|---|
| $a_0$ [ml g$^{-1}$] | 0.7116 | 0.6544 | 0.6168 |
| $a_1$ [ml g$^{-1}$ C$^{-1}$] | 4.68×10$^{-4}$ | 4.04×10$^{-4}$ | 4.30×10$^{-4}$ |
| $a_2$ [ml g$^{-1}$ C$^{-2}$] | 2.8×10$^{-7}$ | 4.7×10$^{-7}$ | 0.7×10$^{-7}$ |
| $b_o$ [MPa] | 229.0 ± 0.7 | 259 ± 1 | 283 ± 4 |
| $b_1$ [C] | -4.89×10$^{-3}$ | -5.24×10$^{-3}$ | -5.20×10$^{-3}$ |
| $\alpha_P$ [C$^{-1}$] [a] | 6.393×10$^{-4}$ | 5.942×10$^{-4}$ | 6.969×10$^{-4}$ |
| $\rho$ [g/mL] [a] | 1.450 | 1.548 | 1.621 |

[a] at $T = T_g$ and $P = 0.1$ MPa

**Table 2** Dynamic properties for supercooled PCB

|  | PCB42 | PCB54 | PCB62 |
|---|---|---|---|
| $T_g$ (K) (DSC) [a] | 227.8 | 249.2 | 268.9 |
| $T(\tau = 10\,s)$ (K) | 224.7 | 251.7 | 273.6 |
| $dT_g/dP$ (K/MPa) | 0.24 ± 0.005 | 0.30 ± 0.005 | 0.31 ± 0.015 |
| $m_\rho$ [b] | 32.9 | 29.0 | 23.4 |
| $\gamma$ | 5.5 | 6.7 | 8.5 |
| $-\alpha_{\tau=10s}/\alpha_{P=0.1MPa}$ | 1.27 | 0.973 | 0.607 |
| $E_V/E_P$ | 0.55$_9$ | 0.49$_6$ | 0.38$_0$ |

[a] measured during cooling at 10 deg/min

[b] at constant $\rho = \rho(T_g, 0.1\,MPa)$

**Figure Captions**

Fig. 1 Selected specific volume data for PCB54 (structure of 3,3'-4,4'-5-pentachlorobiphenyl is shown as representative isomer). Lines through the data are the fits to eq. 2. Vertical tic marks denote the temperature at which $\tau = 10$ s.

Fig. 2 Dielectric relaxation times for PCB54 measured as a function of pressure at T(K) = 283.2 (●), 297.6 (■), 307.75 (▲), and 331.6 (▼). The straight lines are linear fits to the data, yielding the activation volumes displayed in the insert.

Fig. 3 Pressure coefficient of $T_g$ for PCB42 [43], PCB54, and PCB62 [42], we calculate $dT_g/dP = 0.24$ and 0.31 K/MPa respectively, for $\tau(T_g) = 10$ s.

Fig. 4 $T_g$-normalized Arrhenius plots of the PCB relaxation times. The solid symbols are for P = 0.1 MPa, and the hollow symbols for isochoric conditions, at V($T_g$, 0.1 MPa) = 0.6896 (PCB42), 0.6460 (PCB54), and 0.6170 mL/g (PCB62).

Fig. 5 Density-scaled plots of the relaxation times for the PCB, using the indicated values for the exponent. PCB42: P=0.1MPa (○), T=263.2K (△), T=273.2K (☆), T=283.2K (▽); PCB52: P=0.1MPa (○), T=273.2K (□), T=297.6K (△), T=307.8K (▽), T=331.6K (☆); PCB62: P = 0.1MPa (○), T = 295.2K (□), T = 296.2K (△), T = 303.2K (▽), T = 310.2K (◇), T = 314.2K (◁), T = 322.2K (▷), T = 331.2K (○), T = 241.2K (☆).

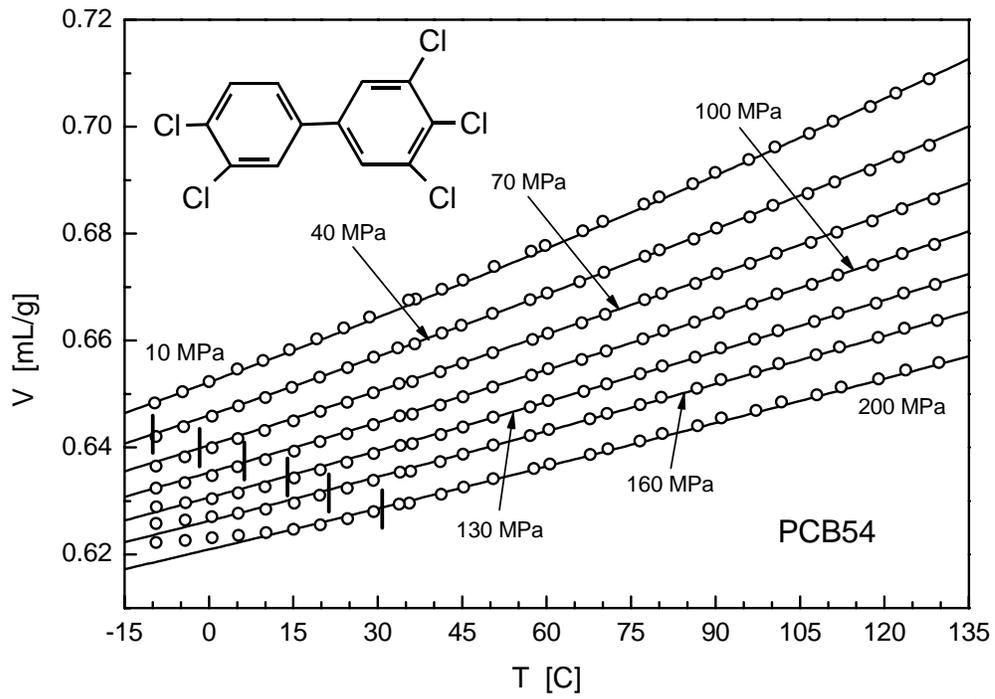

figure 1

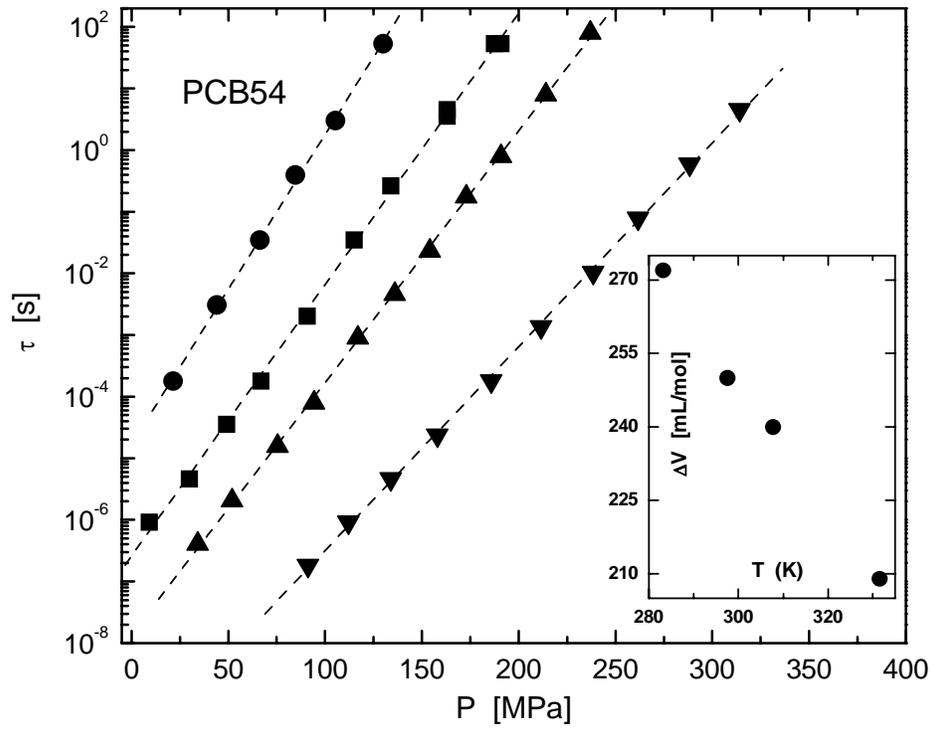

figure 2

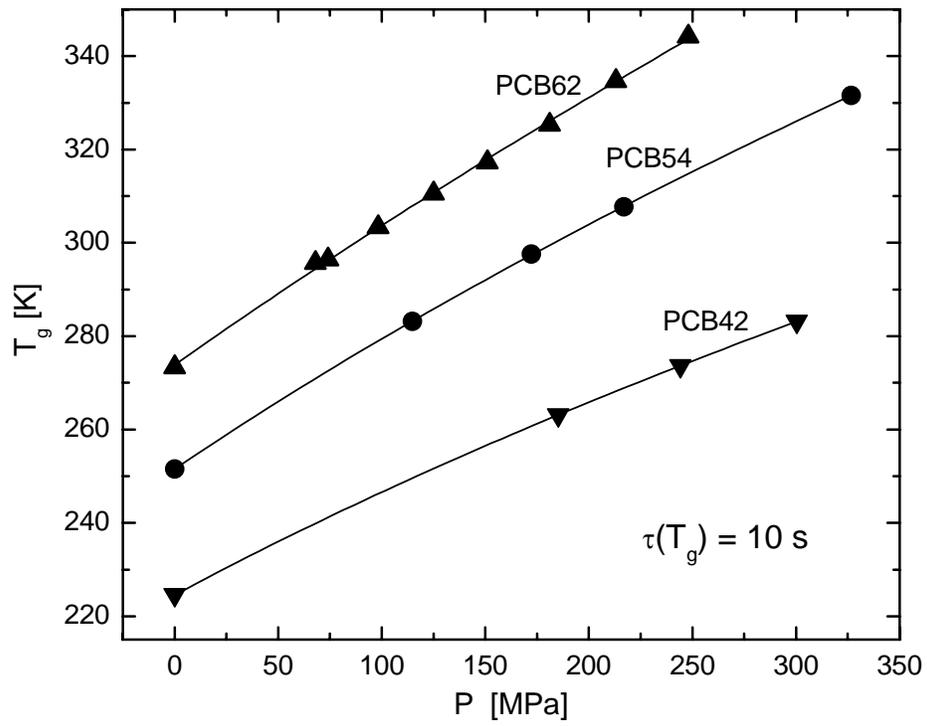

figure 3

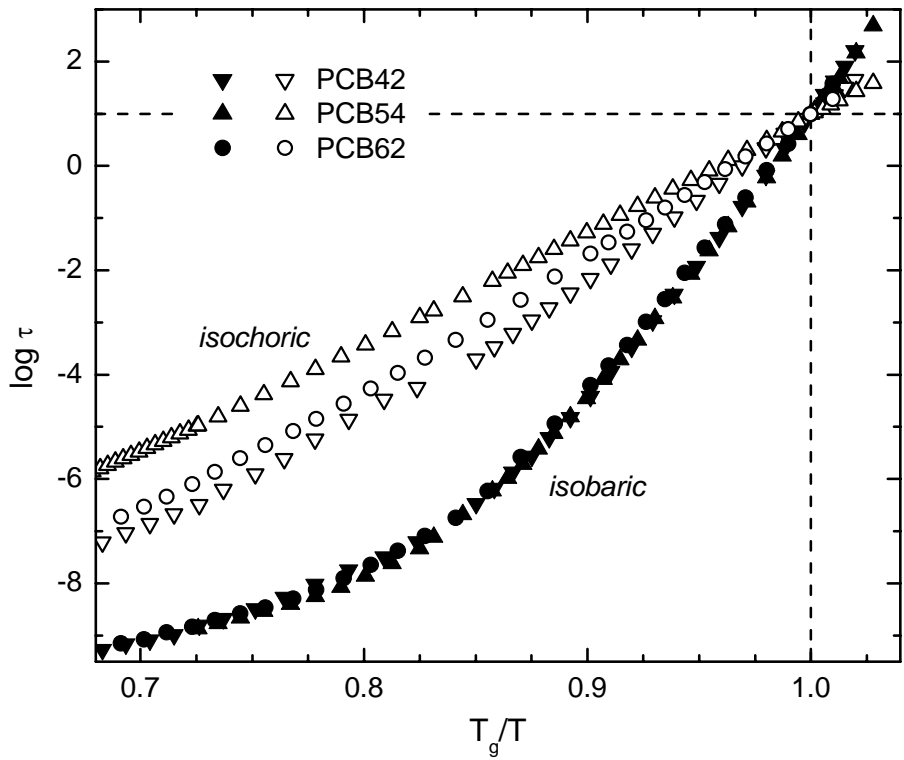

figure 4

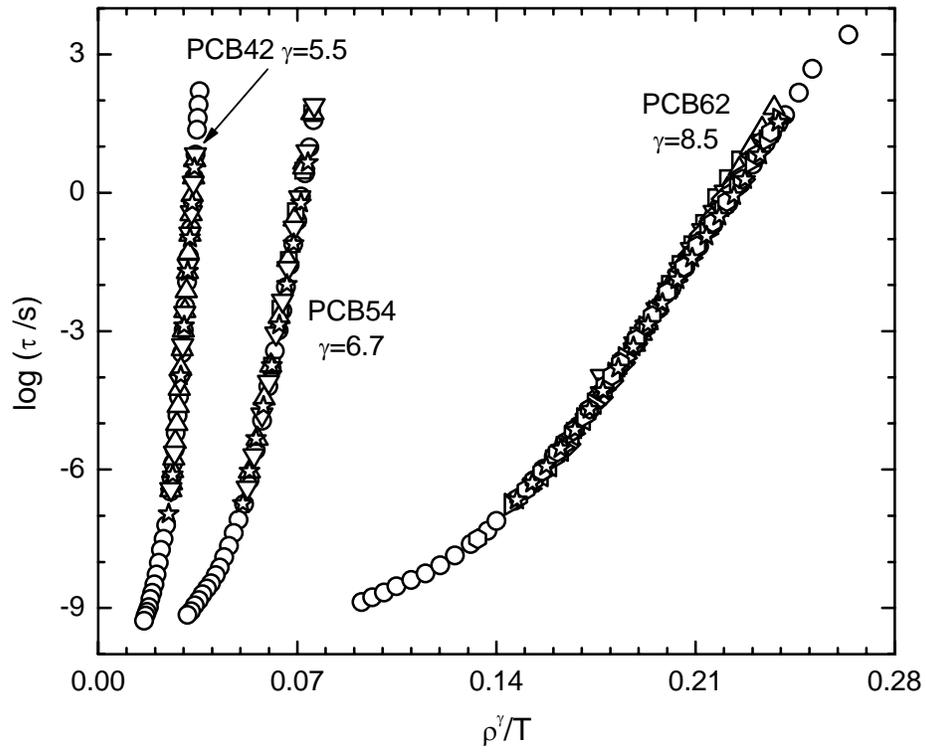

figure 5